\documentclass[3p,times,twocolumn,preprint]{elsarticle}

\usepackage{ecrcJUNK}
\makeatletter

\def\ps@pprintTitle{%
    \def\@evenhead{
      \setbox1=\hbox{\elslogo}%
      \setbox2=\hbox{\sdlogo}%
      \setbox3=\hbox{\jnllogo}%
      \vspace*{2pc}
      \parbox[t]{\wd1}{\elslogo}
       \hfil\parbox[t]{19pc}{\centering%
       \raisebox{23pt}{\sdlogo}\\[-12pt]
       \mbox{\footnotesize\@journalname~(2011)%
         }}\hfil%
        \raisebox{23pt}{\parbox[c]{\wd3}{\jnllogo}}}
      \let\@oddhead\@evenhead%
      \let\@oddfoot\@empty
      \let\@evenfoot\@oddfoot
}

\def\ps@headings{%
    \def\@oddhead{\parbox{\textwidth}{\itshape\footnotesize%
         \hfill\@runauth~\quad~\@journalname~(2011)~%
         \hfill{\rm \thepage}}}%
    \def\@evenhead{\parbox{\textwidth}{\itshape\footnotesize%
         {\rm \thepage}\hfil\@runauth~/~\@journalname~(2011)}}%
    \let\@evenfoot\@empty%
    \let\@oddfoot\@evenfoot}

\makeatother

\def\elslogo{}
\def\sdlogo{}
\def\elslogo{\includegraphics{elsevier-logo-\jtype p}}
\def\sdlogo{\includegraphics{SDlogo-\jtype p}}
\def\elslogo{\includegraphics{elsevier-logo-\jtype p.pdf}}
\def\sdlogo{\includegraphics{SDlogo-\jtype p.pdf}}

\def\elslogo{} \def\sdlogo{}
\volume{00}
\volume{\null}
\volume{2011}

\firstpage{1}

\journalname{Proceedings of the Ringberg Workshop: New Trends in HERA Physics}

\runauth{K.~Kova\v{r}\'ik  et al.}

\jnltitlelogo{Nuclear Physics B Proceedings Supplement}
\jnltitlelogo{Proceedings of the Ringberg Workshop}

\usepackage{amssymb}

\usepackage[figuresright]{rotating}

\begin{document}

\begin{frontmatter}




\title{A Survey of Heavy Quark Theory for PDF Analyses}


\author{
K.~Kova\v{r}\'ik,${}^{a}$
T.~Stavreva,${}^{b}$
A.~Kusina,${}^{c}$
T.~Jezo,${}^{b}$
F.~I.~Olness,${}^{c,1}$
I.~Schienbein,${}^{b}$
J.~Y.~Yu,${}^{b}$
}

\address{${}^{a}$Karlsruhe, Institute of Technology, D-76128, Germany}
\address{${}^{b}$Laboratoire de Physique Subatomique et de Cosmologie, Université
Joseph Fourier/CNRS-IN2P3/INPG, \\
 53 Avenue des Martyrs, 38026 Grenoble, France}
\address{${}^{c}$Southern Methodist University, Dallas, TX 75275, USA}

\begin{abstract}
We survey some of the recent developments in the extraction and application 
of heavy quark Parton Distribution Functions (PDFs). 
We also highlight some of the key HERA measurements 
which have contributed to these advances. 
\end{abstract}


\end{frontmatter}



\addtocounter{footnote}{+1}
\footnotetext{Presented by. F. Olness
at the 
{\it Ringberg Workshop: New Trends in HERA Physics 2011}, 
September 25--28, 2011.
}

\def\gsim{\mathrel{\rlap{\lower4pt\hbox{\hskip1pt$\sim$}} \raise1pt\hbox{$>$}}} 
\def\lsim{\mathrel{\rlap{\lower4pt\hbox{\hskip1pt$\sim$}} \raise1pt\hbox{$<$}}} 
\section*{Two Decades of HERA physics}

\nocite{Kovarik:2010uv,Schienbein:2007fs,Kulagin:2004ie,Kulagin:2007ju,Hirai:2007sx,Bodek:1983qn,Bari:1985ga,Benvenuti:1987az,Landgraf:1991nv,Gomez:1993ri,Dasu:1993vk,Rondio:1993mf,Tzanov:2005kr,Botts:1992yi,Lai:1994bb,Schienbein:2009kk,Aaron:2011gp,Chekanov:2009kj,Breitweg:1999ad,Adloff:2001zj,Kretzer:2003it,Binoth:2010ra,Lai:2010vv,Nadolsky:2008zw,Ball:2011uy,Martin:2010db,Martin:2009iq,Harris:1997zq,Barnett:1976ak,Stavreva:2010xs,Stavreva:2010yh,Abazov:2009de}

\begin{figure}
\includegraphics[clip,width=0.45\textwidth]{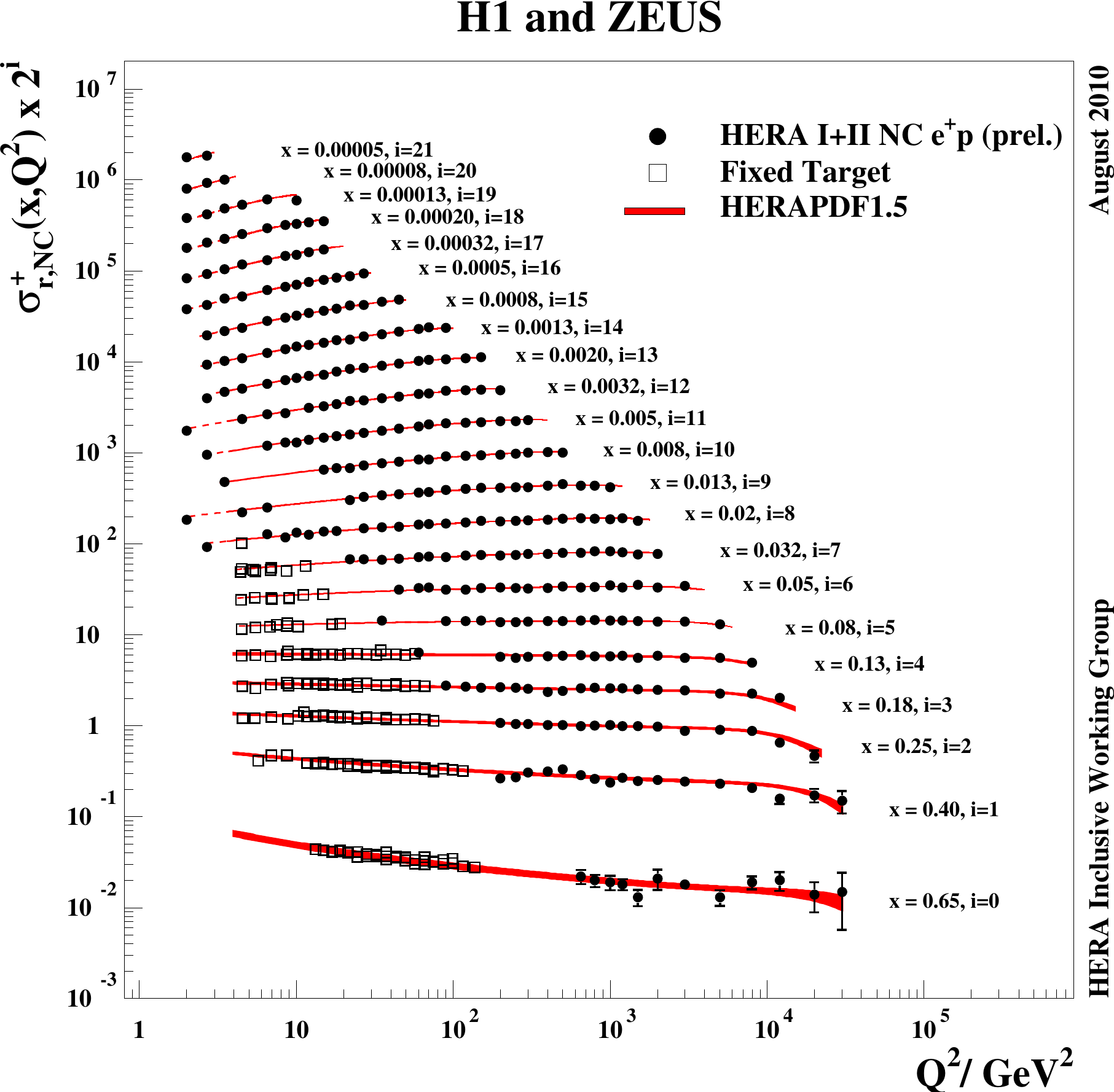}
\caption{
$e^+ p$ NC cross section for the combined HERA data as compared with the 
HERAPDF1.5 fit as a function of $Q^2$ for different values of $x$. 
{\it (Figure  from H1prelim-10-142 \& ZEUS-prel-10-018.)}
\label{fig:herasig}}
\end{figure}

\begin{figure*}[t]
\includegraphics[clip,width=0.45\textwidth]{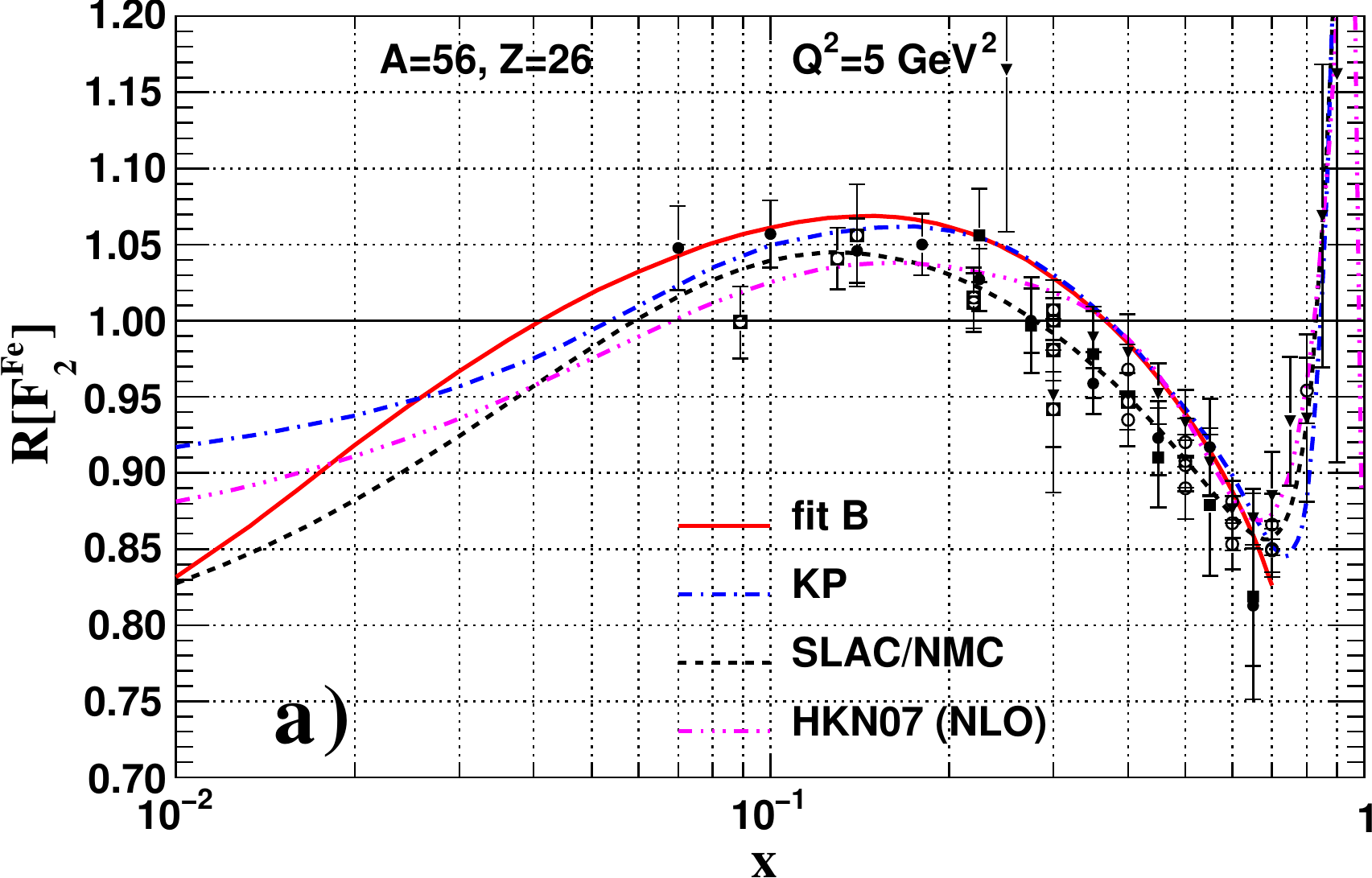}
\hfil
\includegraphics[clip,width=0.45\textwidth]{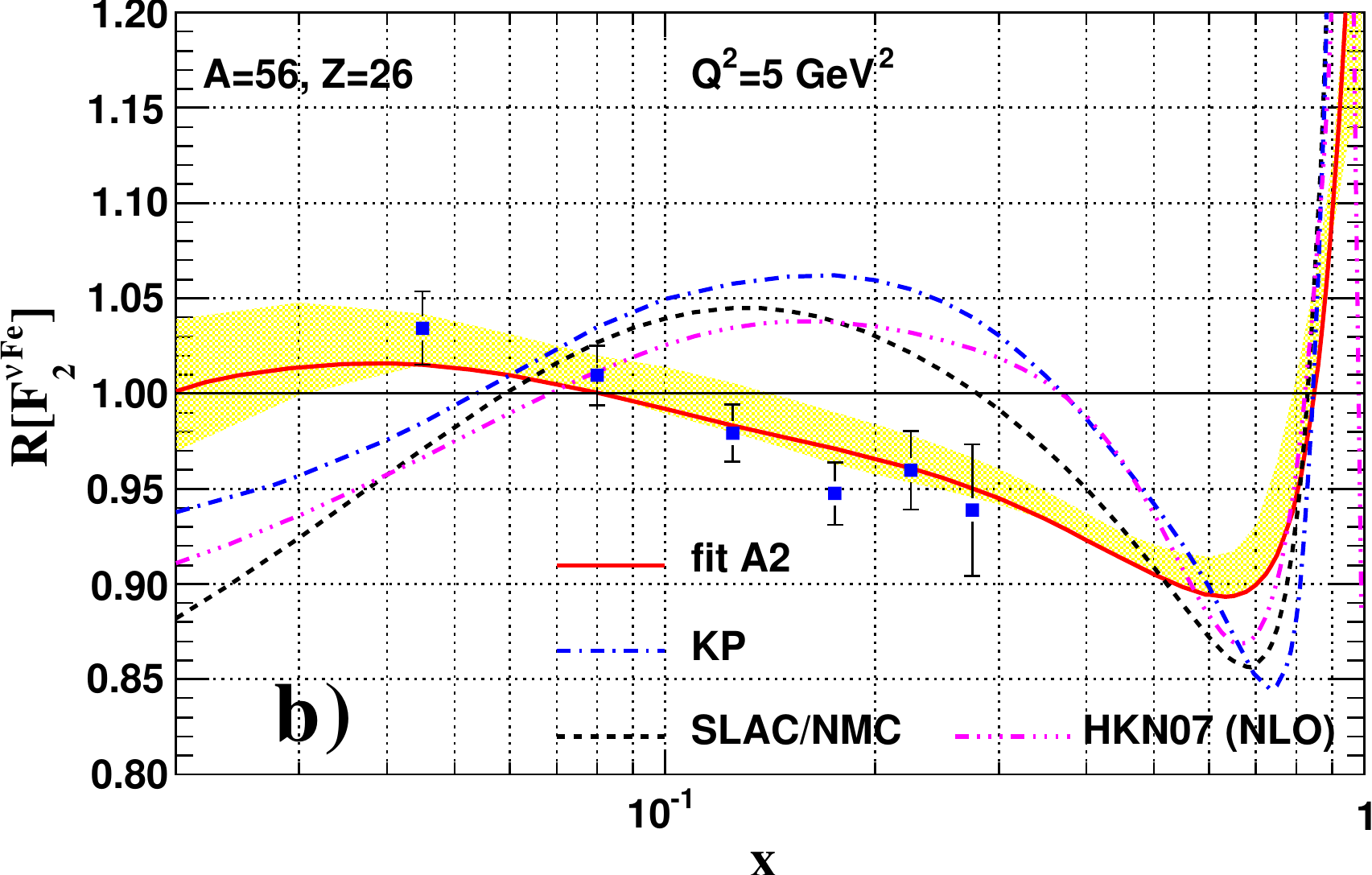}
\caption{The 
computed nuclear correction ratio, $F_{2}^{Fe}/F_{2}^{N}$, as a
function of $x$ for $Q^{2}=5\,{\rm GeV}^{2}$. Figure-a) shows the fit
(fit B from Ref.~\cite{Schienbein:2009kk}) using charged-lepton--nucleus ($\ell^{\pm}A$) and DY data
whereas Figure-b) shows the fit using neutrino-nucleus ($\nu A$) data
(fit A2 from Ref.~\cite{Schienbein:2007fs}). Both fits are compared
with the SLAC/NMC parameterization~\cite{Schienbein:2009kk}, 
 as well as fits from Kulagin-Petti
(KP) (Ref.~\cite{Kulagin:2004ie,Kulagin:2007ju}) and Hirai et
\textit{al.} (HKN07), (Ref.~\cite{Hirai:2007sx}).  The data points
displayed in Figure-a) come from a selection of SLAC and BCDMS
data.~\cite{Bodek:1983qn,Bari:1985ga,Benvenuti:1987az,Landgraf:1991nv,Gomez:1993ri,Dasu:1993vk,Rondio:1993mf}.
\label{fig:nuc}}
\end{figure*}

The HERA electron-proton collider ring began its physics program in
1992 and completed accelerator operations in 2007.
The data collected by the HERA facility allowed for  physics
studies over a tremendously expanded  kinematic region compared to
the previous fixed-target experiments.  This point is illustrated in
Figure~\ref{fig:herasig} where we display the $e^+ p$ Neutral Current
(NC) cross section vs. $Q^2$ for  the HERA data (runs I and II)
together with the fixed-target data.  We observe that the HERA data
allows us to extend our reach in $Q^2$ by more than two decades for
large to intermediate $x$ values, and also extends  the small $x$
region down to $\sim 10^{-5}$.

Additionally, the large statistics and reduced systematics of the
experimental data demand that the theoretical predictions keep pace.
Over the lifetime of HERA we have seen many of the theoretical
calculations advanced from Leading-Order (LO), to
Next-to-Leading-Order (NLO), and some even to
Next-to-Next-to-Leading-Order (NNLO).

As the required theoretical precision has increased, it has been necessary
to revisit the many inputs and assumptions which are used in the calculations.
We will examine the role that the heavy quarks--and
their associated masses--play in these calculations, both 
for the evolution of the parton distribution functions (PDFs) and also the
hard-scattering cross sections.

\section*{Determining the Heavy Quark PDFs}

\begin{figure*}[t]
\includegraphics[clip,width=0.45\textwidth]{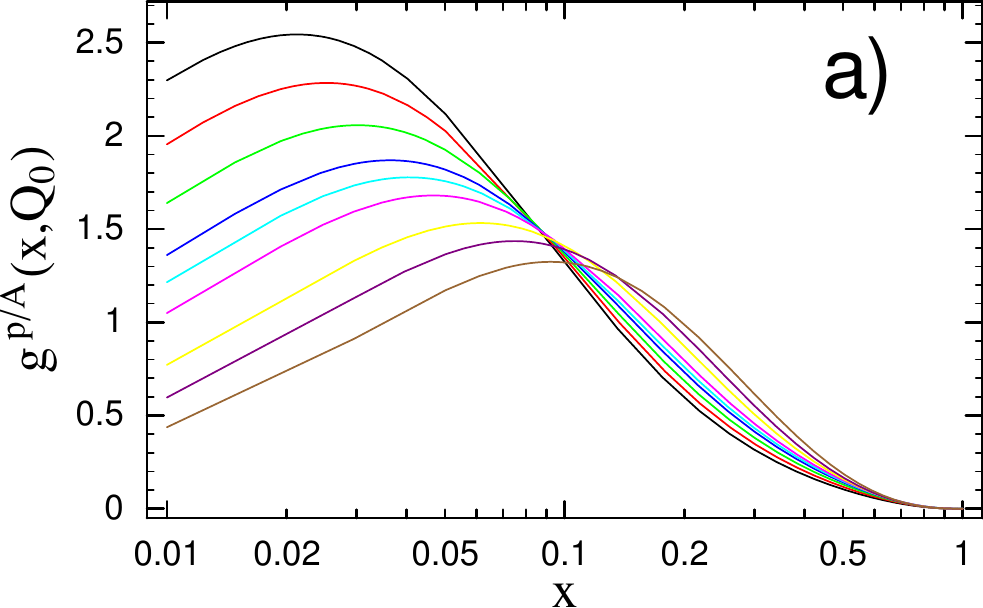}
\hfil
\includegraphics[clip,width=0.45\textwidth]{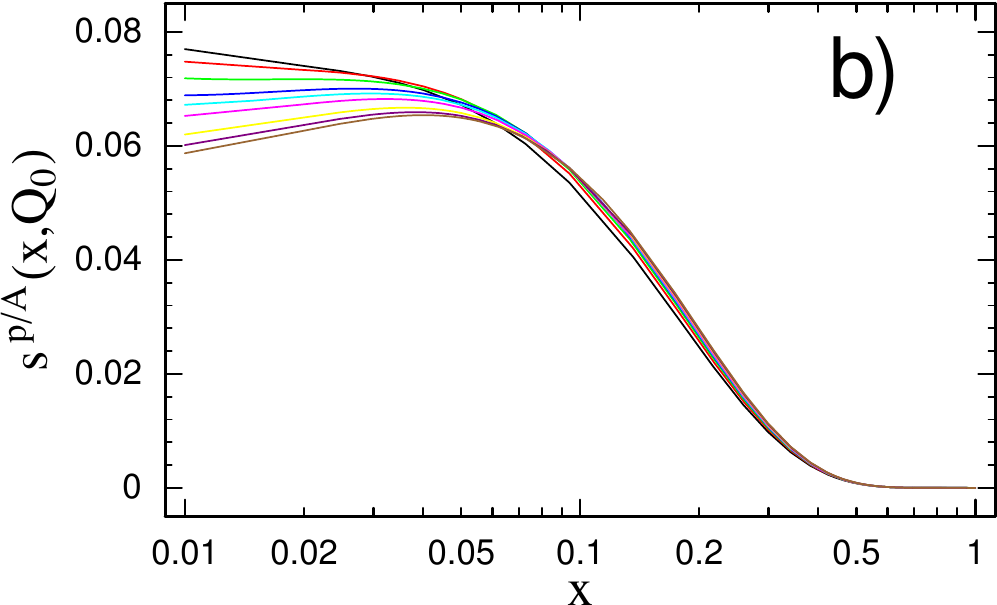}
\caption{The a) gluon $x\, g(x,Q_0)$ and b) strange quark $x\,
s(x,Q_0)$ nuclear PDFs as a function of $x$ for a selection of nuclear
A values $\{1,2,4,9,12,27,56,108, 207 \}$  (from top to bottom at $x=0.01$).
We choose $Q_{0}=1.3\,{\rm
GeV}$.  
\label{fig:npdf}}
\end{figure*}

HERA's reach to larger $Q^2$ and smaller $x$ values takes us to a new
kinematic region where the heavy quarks ($s,c,b$) play a more
important role.  For example, the strange and charm quark
contributions to $F_2$ at small $x$ values can be 30\% or more of the
total inclusive result. To make high-precision predictions for the
structure functions we must therefore be capable of reducing the
uncertainty of these heavy quark contributions; this requires, in part,
precise knowledge of the PDFs which enter the calculation.

The determination of the PDFs requires a variety of data sets which
constrain different linear combinations of the PDF flavors.  For
example, Neutral Current 
(NC) charged-lepton 
Deeply Inelastic Scattering (DIS)
(at low $Q^2$) probes a charge weighted
combination $\sim 4 u + d + s + 4 c $.  In contrast, charged current
(CC) neutrino DIS can probe different flavor combinations via
$W^\pm$-boson exchange; additionally the neutrino measurements can
probe the parity-violating $xF_3$ structure function.

\section*{Nuclear Correction Factors}

The most precise determination of the strange quark PDF component
comes from neutrino--nucleon ($\nu$-N) di-muon DIS  
($\nu N \to \mu^- \mu^+ X$) process. 
This (dominantly) takes place via the Cabibbo favored partonic process 
$\nu s \to \mu^- c$ followed by a semi-leptonic charm decay. 
As the neutrino cross section is small, this measurement 
is typically made using heavy nuclear targets (Fe, Pb), 
so nuclear corrections must be applied to relate the 
results to a proton or isoscalar nuclei. 

However, recent analyses indicate that the nuclear corrections 
for the $\nu$-N and $\ell^\pm$-N DIS processes are different~\cite{Kovarik:2010uv}; 
hence, this introduces an uncertainty into the strange quark PDF extraction 
which was not realized previously. 
Figure~\ref{fig:nuc} displays the nuclear correction factors 
obtained for the 
a) $\ell^\pm$-N 
and 
b)  $\nu$-N processes.
Here, we plot the ratio of $F_2^{Fe}$ to an isoscalar $F_2^N$ 
as a function of $x$ for the $Q^2$ value indicated.

The contrast between the charged-lepton ($\ell^{\pm}A$) case and
the neutrino ($\nu A$) case 
in Figure~\ref{fig:nuc} 
is striking; while the charged-lepton
results generally align with the SLAC/NMC~\cite{Schienbein:2007fs}, 
KP~\cite{Kulagin:2004ie,Kulagin:2007ju} and HKN~\cite{Hirai:2007sx} determinations,
the neutrino results clearly yield different behavior in the intermediate
$x$-region. We emphasize that both the charged-lepton and neutrino
results are not a model---they come directly from global fits to the
data. To emphasize this point, we have superimposed illustrative data
points  in the figures; these
are simply 
a) the SLAC and BCDMS
data~\cite{Bodek:1983qn,Bari:1985ga,Benvenuti:1987az,Landgraf:1991nv,Gomez:1993ri,Dasu:1993vk,Rondio:1993mf}
or
b) the $\nu A$ DIS data~\citep{Tzanov:2005kr}
scaled by the 
appropriate structure function, calculated with the
proton PDF of Ref.~\citep{Schienbein:2007fs}.

The mis-match between the results in charged-lepton and neutrino DIS
is particularly interesting given that there has been a long-standing
{}``tension'' between the light-target charged-lepton data and the
heavy-target neutrino data in the historical fits~\citep{Botts:1992yi,Lai:1994bb}.
This study demonstrates that the tension is not only between charged-lepton
\emph{light-target} data and neutrino heavy-target data, but we now
observe this phenomenon in comparisons between neutrino and charged-lepton
\emph{heavy-target} data.

\section*{The nCTEQ PDFs}

The above example underscores the importance of a comprehensive treatment of the 
nuclear corrections 
to achieve the precision demanded by the current precision data. 
To move toward this goal,  the nCTEQ project was developed to extend the 
global analysis framework of the traditional CTEQ proton PDFs to 
incorporate a  broader set of nuclear data thereby extracting 
the PDFs of a nuclear target. 
In essence, a nuclear  PDF not only depends on the momentum fraction $x$ and 
energy scale $Q$, but  also  on the nuclear ``A'' value: $f(x,Q,A)$.
The structure of the nCTEQ analysis is closely modeled on that of the 
proton global analysis; in fact, the nuclear parameterizations are designed efficiently
to make use of the proton limit (A=1) as a ``boundary condition'' to help constrain the 
fit.

In Figure~\ref{fig:npdf} we display the gluon and strange nuclear PDFs as a function of $x$ 
for a selection of nuclear A values. 
We observe that for $x\simeq 0.01$ the nuclear modifications for the strange quark 
can be $\sim$25\%, 
and for the gluon can be even larger.
The details of the 
nuclear PDF analysis is discussed in 
Refs.~\cite{Kovarik:2010uv,Schienbein:2009kk,Schienbein:2007fs}\footnote{The 
nuclear PDFs are available on the web from the nCTEQ page
at {\tt http://projects.hepforge.org/ncteq/ \ }
which is hosted by the HepForge project.}

The nCTEQ web-page contains 19 families of nPDF grid files which may be used to 
explore the variation due to the different data sets and kinematic cuts.  In particular, 
there is a collections of nPDFs which interpolate between that
of Figure~\ref{fig:nuc}-a) which uses the charged-lepton--nucleus ($\ell^{\pm}A$) 
data and 
of Figure~\ref{fig:nuc}-b) which uses the  neutrino-nucleus ($\nu A$) data.

\section*{The Strange Quark PDF}

\begin{figure}
\centerline{
\includegraphics[clip,width=0.40\textwidth]{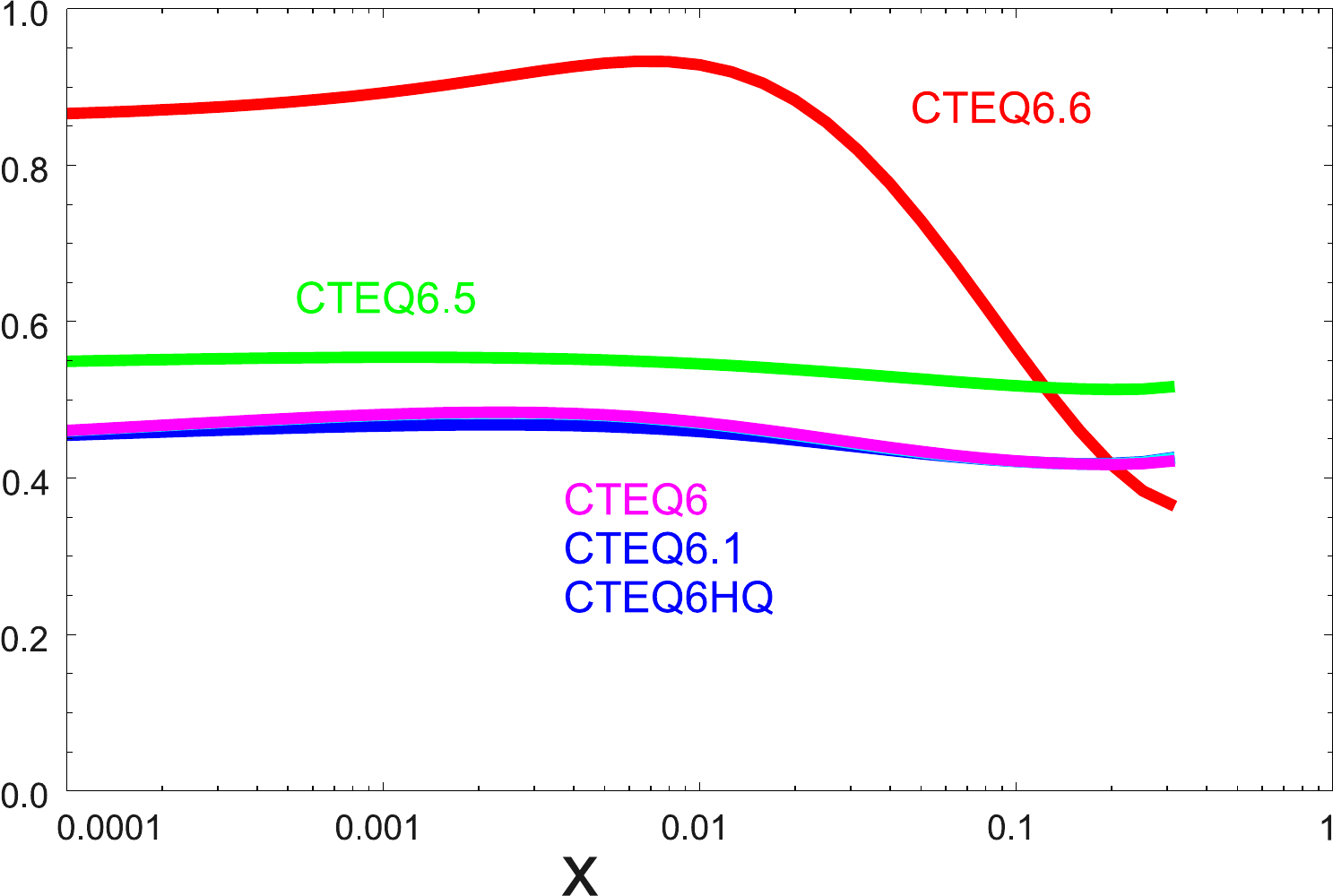}
}
\caption{$\kappa(x)$ vs. $x$ for $Q=1.5$~GeV 
for a selection of PDFs, where we define 
$\kappa(x)=2 s(x)/(\bar{u}(x)+\bar{d}(x))$ 
\label{fig:kappa}}
\end{figure}

\begin{figure}
\centerline{
\includegraphics[clip,width=0.40\textwidth]{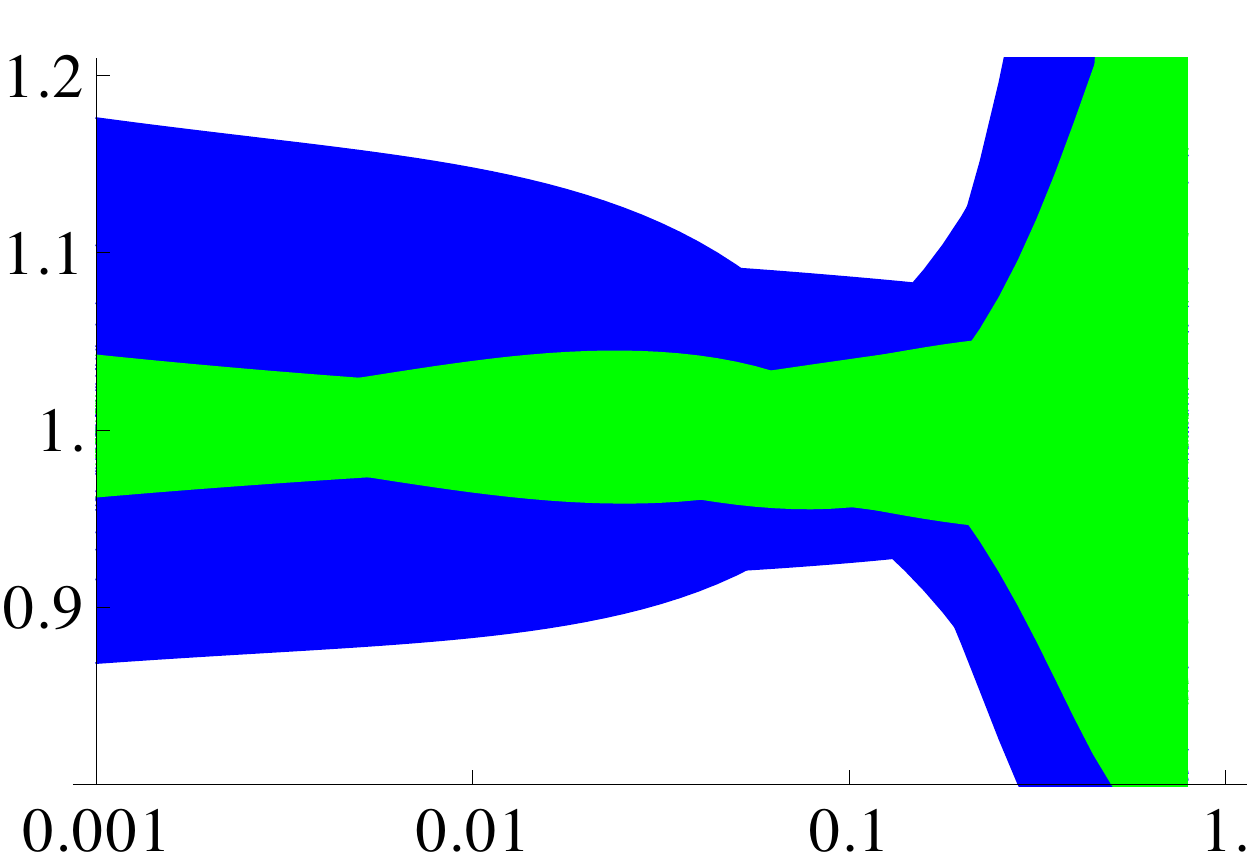}
}
\caption{Relative uncertainty of the strange quark PDF 
as a function of $x$ for $Q=2$~GeV.
The inner band is for the CTEQ6.1 PDF set, and the 
outer band is for the CTEQ6.6 PDF set. 
The band is computed as 
$s_i(x)/s_0(x)$ where $s_0(x)$ is the central PDF for each set;
for CTEQ6.1, $i=[1,40]$, and 
for CTEQ6.6, $i=[1,44]$.
\label{fig:band}}
\end{figure}

We now 
compare a selection of $s(x)$ distributions to gain 
a better understanding of the uncertainties
arising from the nuclear correction factors used to analyze the  
$\nu$N DIS.
One measure of the strange quark content of the proton is to compare 
$s(x)$ with the average up-quark and down-quark sea PDFs: 
$(\bar{u}(x)+\bar{d}(x))/2$. 
Thus, we define the ratio 
$\kappa(x)=2 s(x)/(\bar{u}(x)+\bar{d}(x))$.
If we had exact  $SU(3)$ flavor symmetry, we would expect $\kappa=1$; 
the extent to which $\kappa$ is below one  measures  the suppression of the
strange quark as compared to the up and down sea. 
In Figure~\ref{fig:kappa} we display $\kappa(x)$ for some recent CTEQ PDFs and 
note that $\kappa(x)$ has a large variation, especially at small $x$ values.
This reflects, in part, the fact that the strange quark is poorly constrained 
for $x\lsim 0.1$. 
For the CTEQ6, 6HQ, 6.1, and 6.5 PDF sets, 
the strange quark was arbitrarily set to $\sim 1/2$ the 
average of the up and down sea-quarks. 
For the CTEQ6.6 PDF set, the strange quark was allowed 
additional freedom;  this 
is reflected in Figure~\ref{fig:band}
which compares the relative uncertainty of the strange quark
in the 6.1 and 6.6 PDF sets.

\section*{$W/Z$ at the LHC}

\begin{figure}
\centerline{
\includegraphics[clip,width=0.40\textwidth]{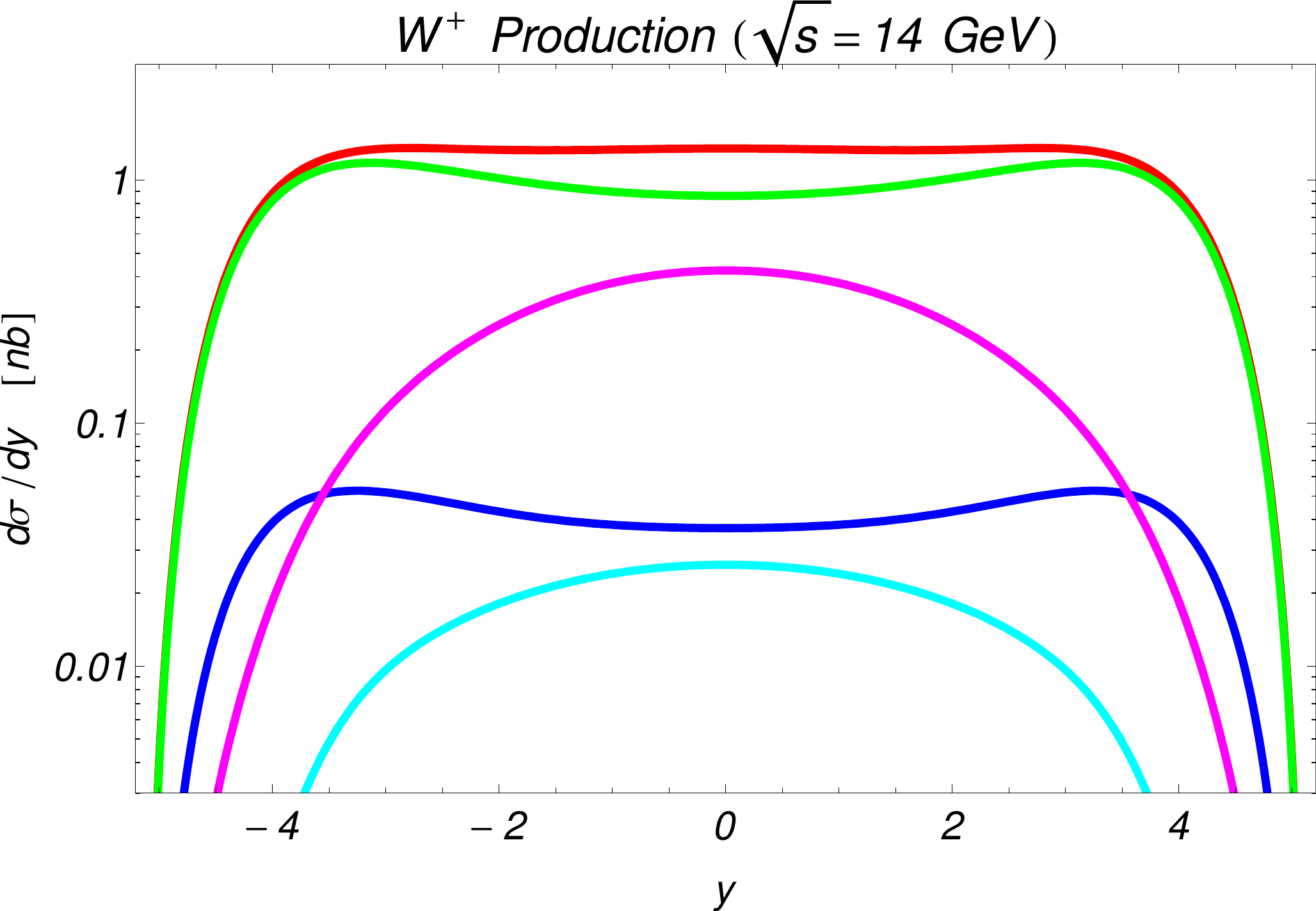}
}
\caption{The differential cross section ($d\sigma/dy$) for $W^+$ production 
at the LHC ($\sqrt{s}$=14TeV) as a function of rapidity $y$. 
The partonic contributions are also displayed. At $y=0$, the contributions 
(from top to bottom) are 
\{$total$, 
$u \bar{d}$,
$c \bar{s}$,
$u \bar{s}$,
$c \bar{d}$ \}
\label{fig:lhc}}
\end{figure}

The above reexamination of the nuclear corrections introduces additional 
uncertainties into the data sets, which manifests itself in increased uncertainties 
on the strange quark PDF. To see how these uncertainties might affect other processes, 
we consider, as an example, $W/Z$ production at the LHC. 
As we go to higher energies,
the heavy quarks will play an increasingly important role
because we can probe the PDFs at smaller $x$ and larger $Q$;
this means that 
the heavy quark PDF uncertainties can 
have an increased influence on LHC observables compared to Tevatron observables.

In Figure~\ref{fig:lhc} we display the LO differential cross section for 
$W$ production at the LHC as a function of rapidity $y$, as well as the 
individual partonic contributions. We note that in the central rapidity region
the contribution from the heavy quarks can be  30\% or more of the total cross section; 
this is in sharp contrast to the situation at the Tevatron where the 
heavy quark contributions are minimal. Thus, a large uncertainty in the 
heavy quark PDFs can influence such ``benchmark'' processes as W/Z production 
at the LHC. Of course, given the high statistics from the LHC (the 2011 proton-proton run exceeded 5~$fb^{-1}$), 
it may be possible to turn the question around and ask to what extent the 
LHC data may constrain the heavy quark PDFs.

\section*{Zero Mass (ZM) and General Mass (GM) Schemes}

We now turn to charm production and the measurement of the charm PDF.
HERA extracted precise measurements of $F_2^{c}$ and $F_2^{b}$, and recently these
analyses have been updated\footnote{Cf., ZEUS-prel-09-015}
to include the low $Q^2$ data to cover the
kinematic range of $Q^2=[2, 1000] \, GeV^2$ and $x$ down to $10^{-5}$.
~\cite{Aaron:2011gp,Chekanov:2009kj}

A  global fit of HERA I data~\cite{Breitweg:1999ad,Adloff:2001zj}
 for  $F_2^{c}$ was performed using both the 
General Mass Variable Flavor Scheme (GM-VFS), and also the 
Zero Mass Variable Flavor Scheme (ZM-VFS).~\cite{Kretzer:2003it}
While the GM-VFN result yielded an improved $\chi^2$, 
the ZM-VFN results--when implemented consistently--yielded an acceptable fit to the data. 
Given the expanded kinematic coverage of the recent HERA data, it would be of interest to 
repeat this comparison. 
Presumably the new data sets would allow for increased differentiation 
between the ZM-VFN and GM-VFN scheme results.

\section*{Choice of Theoretical Schemes}

\begin{figure}
\includegraphics[clip,width=0.40\textwidth]{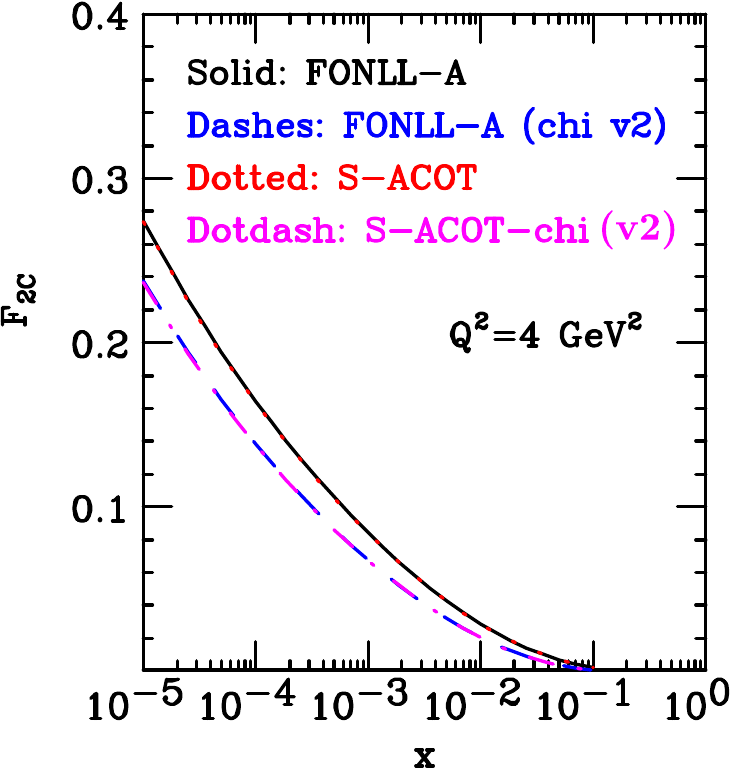}
\caption{Comparison of  $F_2^c$ for the 
Fixed-Order-Next-to-Leading-Log (FONLL) with the 
Simplified-ACOT (S-ACOT) scheme. 
There are four curves displayed, two for the 
ordinary $\chi$-rescaling, and two for 
an alternate  $\chi$-rescaling (labeled ``v2''). 
The FONLL  and S-ACOT results are identical throughout 
the $x$ range. 
{\it (Figure  from Ref.~\cite{Binoth:2010ra}.)}
\label{fig:lh1}}
\end{figure}

\begin{figure}
\includegraphics[clip,width=0.40\textwidth]{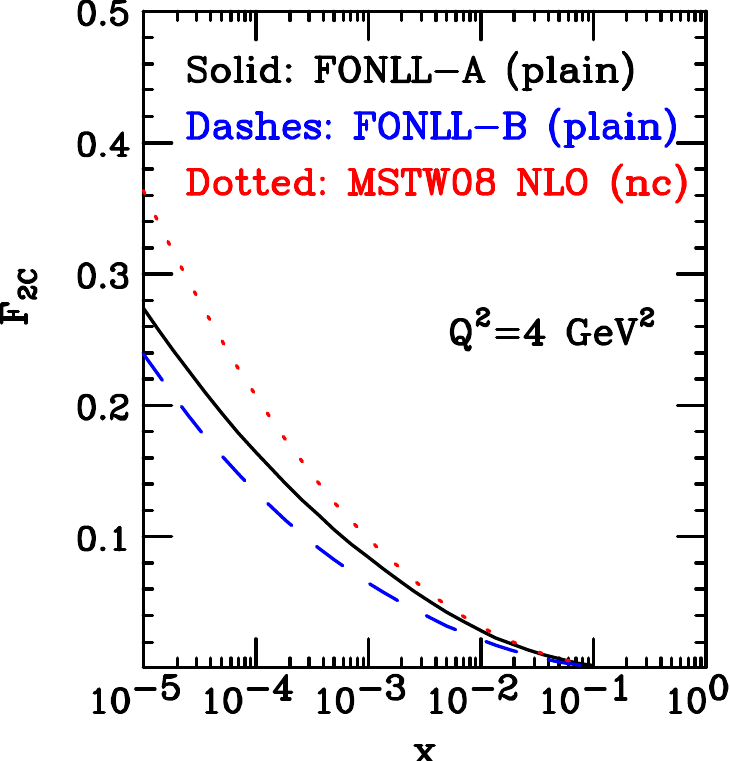}
\caption{Comparison of  $F_2^c$ for the 
Fixed-Order-Next-to-Leading-Log (FONLL) and the 
MSTW08 NLO results. 
The FONLL results are shown for both the ``A'' and ``B'' 
variations. 
The  FONLL results differ from the MSTW08 results 
for low $Q$ and $x$; for larger values of $Q$ and $x$ 
they are more comparable. 
{\it (Figure  from Ref.~\cite{Binoth:2010ra}.)}
\label{fig:lh2}}
\end{figure}

Having illustrated the impact of different theoretical schemes 
on the data analysis, we take a moment to compare and contrast 
some of the different schemes that are currently being used 
for various PDF analysis efforts. 
While many of the global analyses use a 
Variable Flavor Number (VFN) scheme to include the
heavy quark as a parton, the detailed implementation of this scheme
can lead to notable differences.  At the 2009 Les Houches workshop, a
comparison was performed among a number of the different programs to
quantify these differences.  All programs used the same PDFs and
$\alpha_S$ values so that the differences would  only reflect the
particular scheme.  The complete details can be found in
Ref.~\cite{Binoth:2010ra}, and Figures~\ref{fig:lh1} and
\ref{fig:lh2} display sample comparisons.

Figure~\ref{fig:lh1} compares the S-ACOT scheme which is used for the
CTEQ series of global analyses,~\cite{Lai:2010vv,Nadolsky:2008zw}
 and the FONLL which is used by the
Neural Network PDF (NNPDF)
collaboration.~\cite{Ball:2011uy}
In the figure, these two implementations (S-ACOT and FONLL-A) are numerically equivalent.

Figure~\ref{fig:lh2} compares two variations of the  FONLL scheme (``A'' and ``B'')  
with the MSTW08 results which is used in the MSTW series of PDF global analyses.~\cite{Martin:2010db,Martin:2009iq}
Here these differences reflect the different organization 
and truncation of the perturbation expansion; it does {\it not} indicate that
one choice is right or wrong. 
We expect such differences to be proportional to $\sim \alpha_S^N \times {\cal O}(m^2/Q^2)$. 
Thus, 
as we increase the order of perturbation theory  or the energy scale  
the differences should decrease; 
we have explicitly verified the difference is reduced as $Q^2$ increases, as it should. 
When we are able to carry these calculations out to higher orders, the scheme differences 
should be further reduced; this work is in progress.

\section*{Charm Mass Dependence and $F_2^{c}$}

The experimental extraction of the ``inclusive'' $F_2^{c}$ requires a
differential NLO calculation of DIS charm production to be extrapolated
over the unobserved kinematic regions.
These analyses generally make use of the HVQDIS
program~\cite{Harris:1997zq} which computes $F_2^{c}$ in a
Fixed-Flavor-Number  (FFN) scheme at NLO. 
In this calculation, the charm is produced only
via a gluon splitting, $g\to c \bar{c}$, and there is no charm PDF.
Thus, the charm mass ($m_c$) enters only the partonic cross section $\hat{\sigma}(m_c)$ and the 
final state phase space; there is no PDF charm threshold. 

Although we would also like to perform the extraction of $F_2^{c}$ 
using a Variable-Flavor-Number (VFN) scheme, the challenge is that no NLO differential 
program exists for this process. 
In lieu of a VFN extraction, another avenue is to study the  
influence of different theoretical schemes and $m_c$ parameters in the analysis of 
the  $F_2^{c}$ data. We describe such a study below.

%

\begin{figure}[t]
\includegraphics[clip,width=0.45\textwidth]{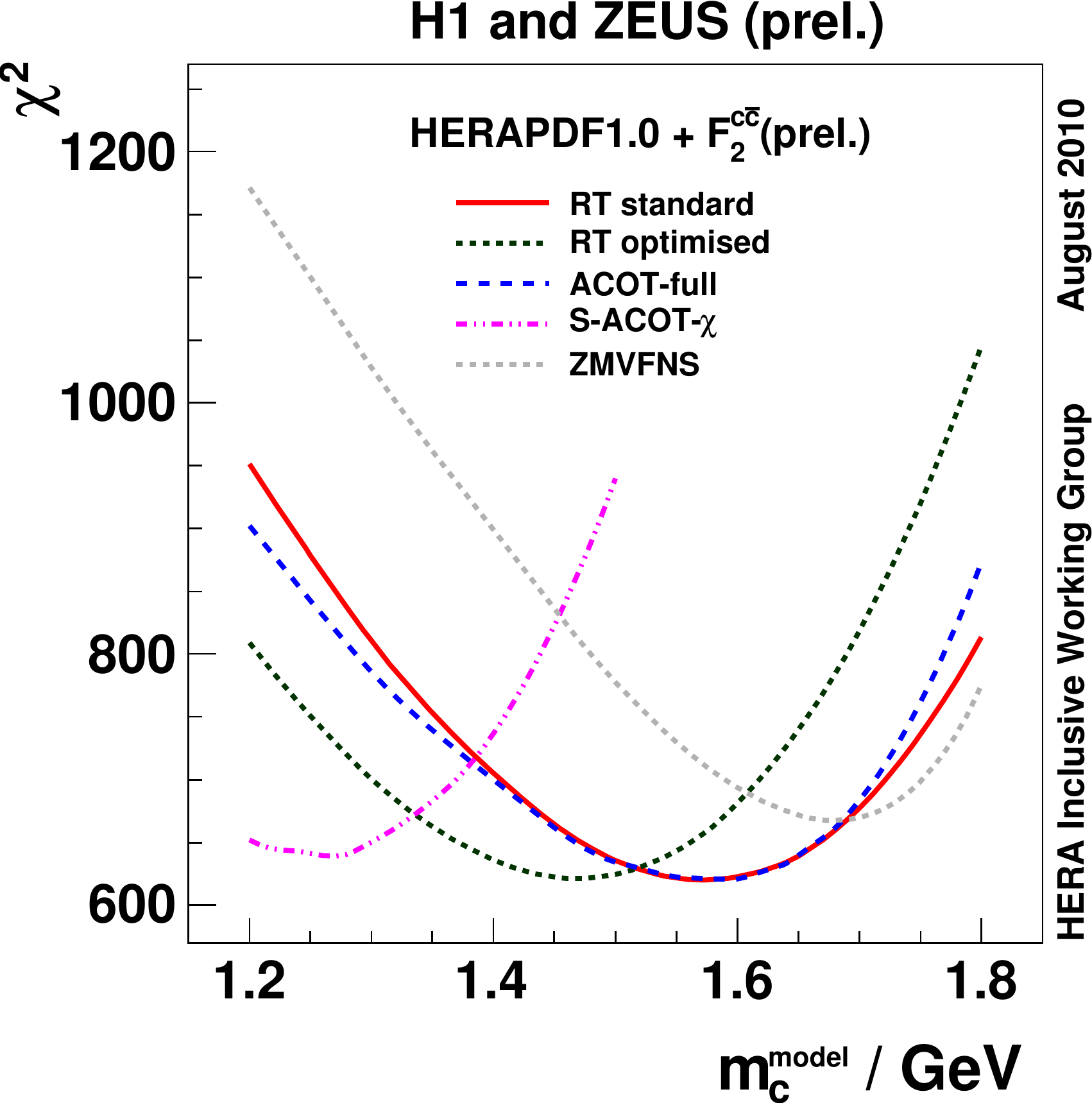}
\caption{Comparison of $\chi^2$ for  HERA I + $F_2^{c \bar{c}}$ fits 
using different heavy flavor schemes as a function of the 
charm quark mass parameter $m_c^{model}$.
{\it (Figure  from H1prelim-10-143 \& ZEUS-prel-10-019)}
\label{fig:charm}}
\end{figure}

In many analyses, the value of the charm mass is taken as an external
fixed parameter.
A recent investigation has taken a closer look at the role of the
charm mass parameter $m_c$ and examined the combined effects of $m_c$
and the theoretical scheme used; preliminary results 
of this study are displayed in
Figure~\ref{fig:charm}. 
For each of the schemes listed in the legend, fits were generated for
fixed $m_c$ values in the range $[1.2, 1.8]$~GeV.  Thus, the minimum
of the $\chi^2$ curve represents the ``optimal'' choice of the charm
mass parameter $m_c$ for that specific scheme.

We observe that the various schemes prefer $m_c$ values ranging from
1.2 to 1.7~GeV,
The largest  $m_c$ value (1.68~GeV) comes from the Zero Mass VFN Scheme (ZM\, VFNS) 
which only uses $m_c$ for the PDF charm threshold; it is absent in the 
 phase space for the zero-mass case. 
In contrast, for the S-ACOT-$\chi$ scheme  the ``$\chi$'' notation\footnote{Specifically, 
the $\chi$-prescription rescales the partonic momentum fraction via 
$x \to x(1+(2m_c/Q)^2)$
in contrast to the traditional Barnett~\cite{Barnett:1976ak} ``slow-rescaling''
which is  $x \to x(1+(m_c/Q)^2)$.}
 indicates
there are effectively two factors of $m_c$ in the final phase space, and this 
yields the smallest value of $m_c$ (1.26~GeV).

The ACOT scheme and the Roberts-Thorne (RT) scheme yield $m_c$ values 
in the intermediate region.
The ACOT scheme uses the full kinematic mass relations in the partonic relations, 
and the scaling variable is intermediate ($m_c$=1.58~GeV)
between the ZM-VFN scheme and the S-ACOT-$\chi$ scheme. 
The S-ACOT scheme (not shown) is virtually identical to the full ACOT scheme, and also 
yields a $m_c$ value  in the intermediate region. 
Therefore, comparing the  S-ACOT-$\chi$ and S-ACOT schemes, we find that the 
$\chi$-rescaling variable 
is dominantly responsible for the shift of the optimal $m_c$ value from $\sim 1.26$ to 
 $\sim 1.58$.
This observation suggests that it is the rescaling of the $x$ variable which enters the 
PDFs that generates  the dominant effect.
While this study is continuing, it does indicate 
the sensitivity  of the charm mass and scheme choice 
in these precision analyses.

\section*{Extrinsic \& Intrinsic Charm PDFs}

\begin{figure}[t]
\includegraphics[clip,width=0.45\textwidth]{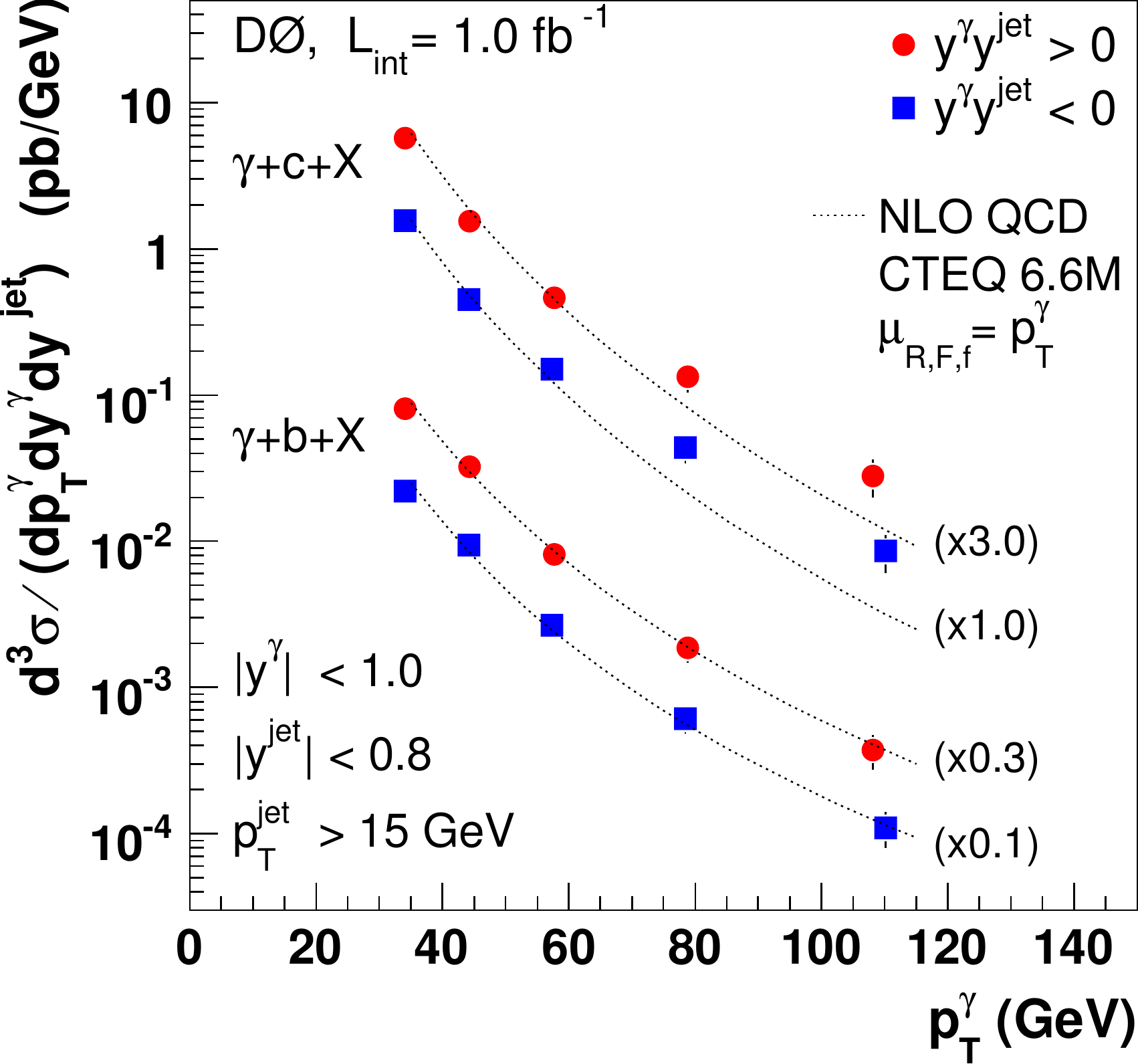}
\caption{Measurement of $\gamma+c$ and $\gamma+b$ vs. 
$P_T^\gamma$ as measured by D-Zero.
{\it (Figure  from Ref.~\protect\cite{Abazov:2009de}.)}
\label{fig:d0}}
\end{figure}

\begin{figure}[t]
\hfil
\includegraphics[clip,width=0.30\textwidth]{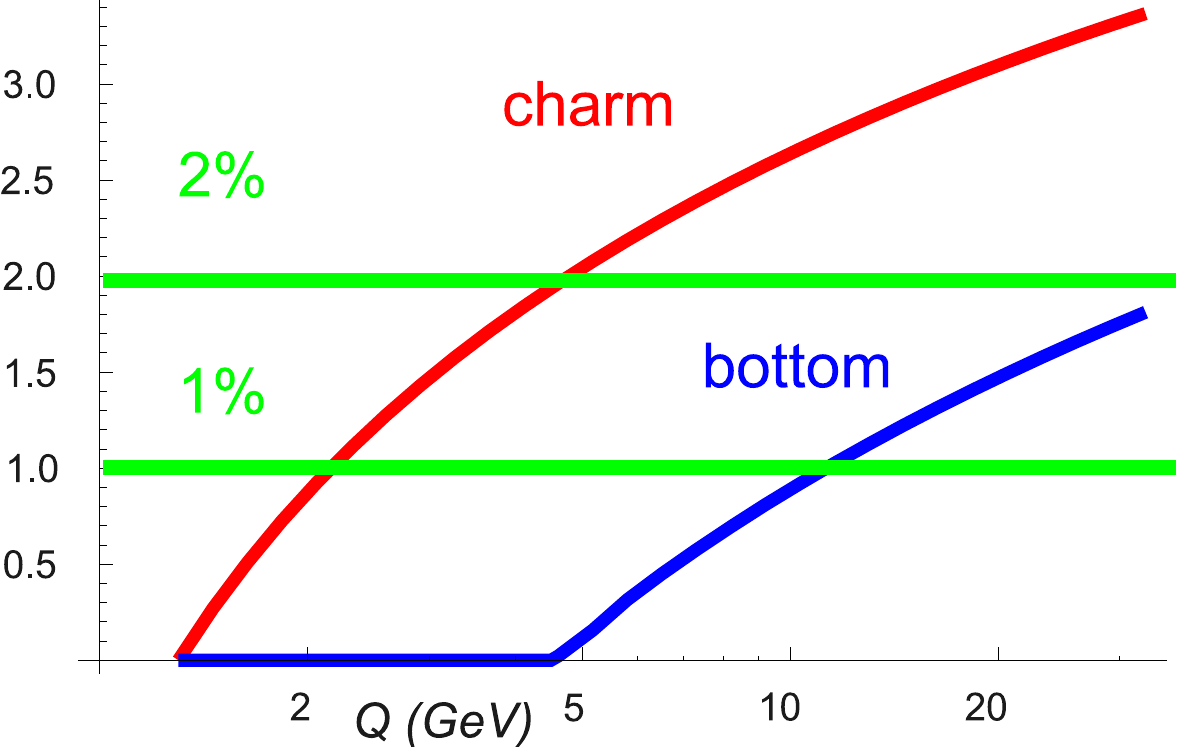}
\hfil
\caption{The integrated momentum fraction (in percent) of the ``extrinsic'' 
charm and bottom quarks generated by gluon splitting 
as a function of the scale $Q$. Reference lines are indicated at 1\% and 2\%. 
\label{fig:mom}}
\end{figure}

\begin{figure}[t]
\includegraphics[clip,angle=0,width=0.45\textwidth]{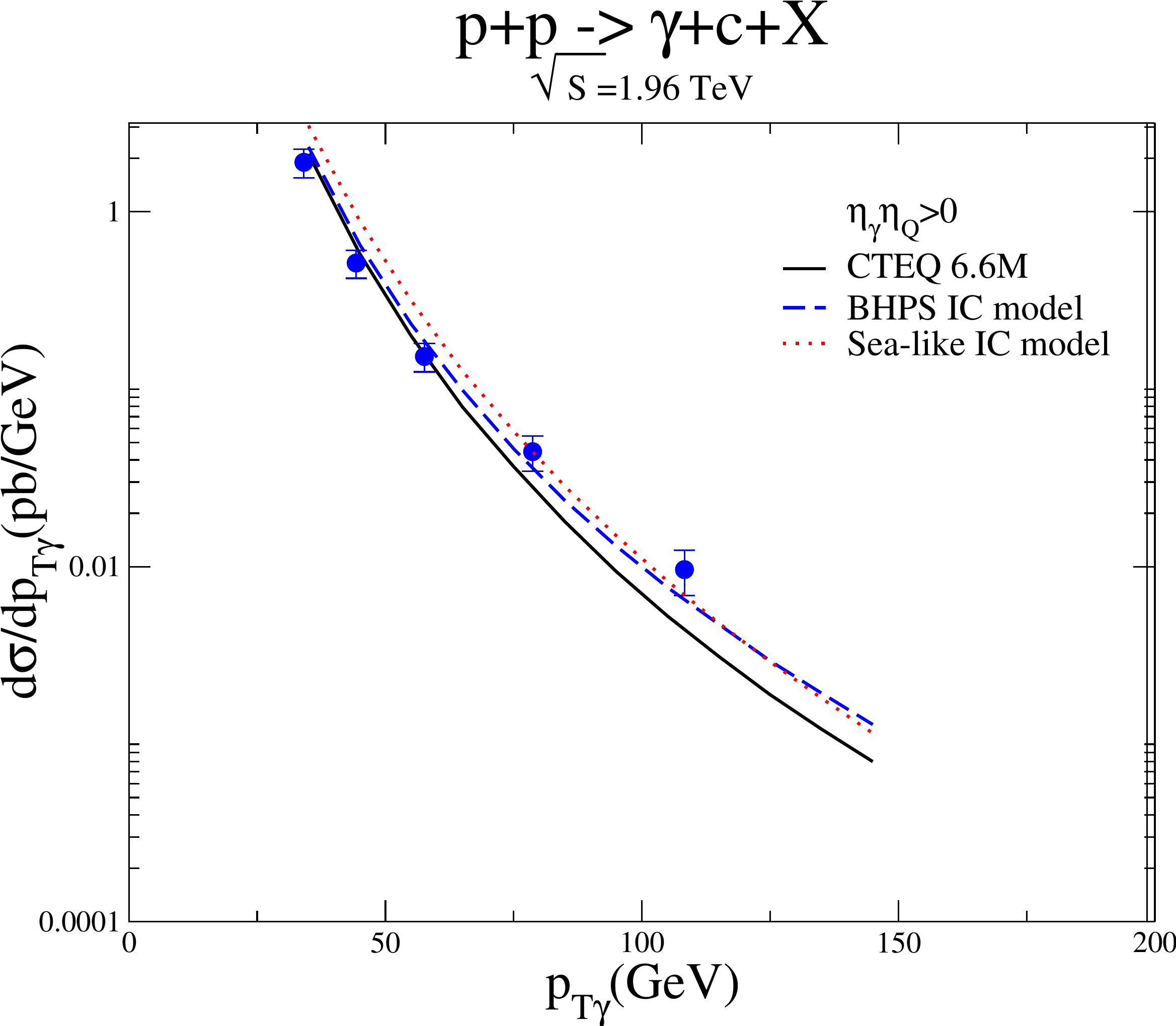}
\caption{$d\sigma/dp_{T\gamma}$ (pb/GeV) 
at the Tevatron for the CTEQ6.6 PDFs, and two 
intrinsic charm (IC) models. 
{\it (Figure  from Ref.~\cite{Stavreva:2010xs}.)}
\label{fig:charmTEV}}
\end{figure}

\begin{figure}[t]
\includegraphics[clip,angle=0,width=0.45\textwidth]{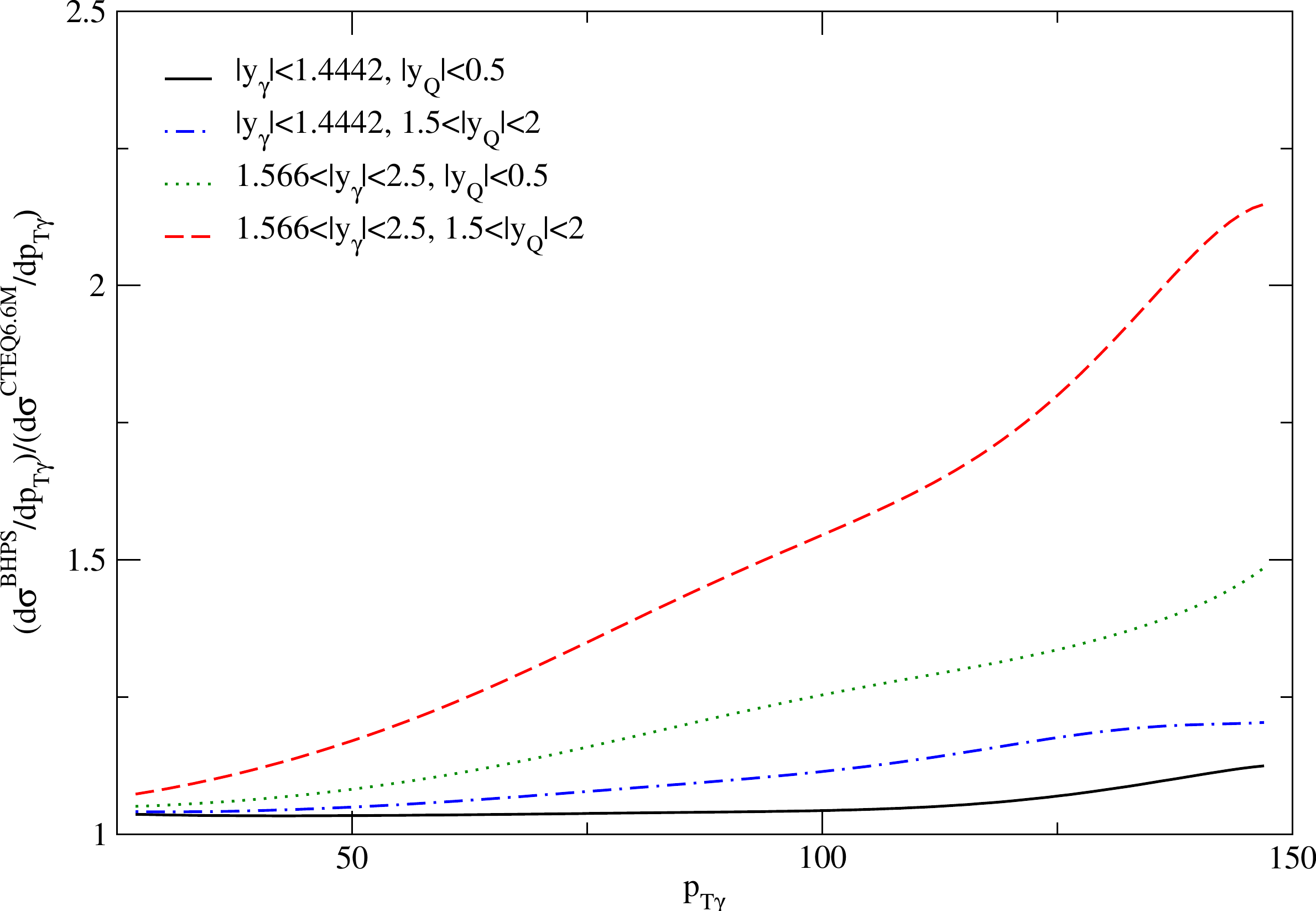}
\caption{Ratio of $d\sigma/dp_{T\gamma}$ (pb/GeV) 
at the LHC ($\sqrt{s}=7~TeV$) for the 
 BPHS IC model  to the 
CTEQ6.6 PDF 
for a selection of rapidity bins. 
{\it (Figure  from Ref.~\cite{Stavreva:2010xs}.)}
\label{fig:charmLHC}}
\end{figure}

The charm quark and bottom quark PDFs can  be probed directly 
at the Tevatron by studying photon--heavy quark final states 
which occur via the sub-process  $g Q \to \gamma Q$ at LO. 
This process has been measured at the Tevatron for both charm and bottom
final states, 
and we display the results in Figure~\ref{fig:d0} as a function 
of $P_T^\gamma$ for two rapidity configurations. 
This measurement is particularly interesting as the dominant process involves a 
heavy quark PDF; this is in contrast to DIS charm or bottom production, for example, where 
over much of the kinematic range the process is dominated by the gluon-initiated process
(e.g., $\gamma g \to Q \bar{Q}$) rather than the heavy quark initiated process ($\gamma Q \to Q$).

Examining Figure~\ref{fig:d0} we observe that 
the bottom quark production measurements compare favorably with the theoretical predictions 
throughout the  $P_T^\gamma$ range, but the charm results rise above the theory predictions for 
large  $P_T^\gamma$.
Although there may be a number of explanations for the excess charm
cross section at large $P_T^\gamma$, one possibility is the presence
of intrinsic charm (IC) in the proton.  In the usual DGLAP evolution of the
proton PDFs, we begin the evolution at a low energy scale $Q_0<m_c$
and evolve up to higher scales.  The charm and bottom PDFs are defined
to be zero for $Q<m_{c,b}$, and above the mass scale the heavy quark
PDFs are generated by gluon splitting, $g\to Q \bar{Q}$; we refer to
this as the ``extrinsic'' contribution to the heavy quark PDFs. 
In Figure~\ref{fig:mom} we display the integrated momentum fractions for the 
charm and bottom quarks as a function of the scale $Q$; these are zero for 
$Q<m_{c,b}$, and then begin to grow via the  $g \to Q \bar{Q}$ process.

It has been suggested that there may also be an ``intrinsic
contribution to the heavy quark PDFs which is present even at low
scales $Q<m_{c,b}$.
While it is difficult to constrain the detailed functional shape of
any intrinsic heavy quark distribution, the total momentum fraction of
any intrinsic contribution must be less than  approximately $1\%$ if it is to be
compatible with the global analyses.

In Figure~\ref{fig:charmTEV} we illustrate the effect of including 
an additional intrinsic charm component in the proton. 
The BHPS IC model concentrates the momentum fraction at large $x$ values, 
and the Sea-like IC model distributes the charm more uniformly.\footnote{For 
details, c.f., Refs.~\cite{Stavreva:2010xs,Stavreva:2010yh}}
It is intriguing that the IC modification 
of the proton PDF can increase the theoretical prediction in the large  $P_T^\gamma$ region,
but this observation alone is not sufficient to claim the presence of IC; this would require
independent verification. 
In Figure~\ref{fig:charmLHC} we display the cross section ratio for $\gamma + c$
at the LHC for the BHPS IC model for a selection of rapidity bins. 
Thus the LHC can validate or refute this possibility with 
a high-statistics measurement of $\gamma + c$, especially if they can observe this in the 
forward rapidity region.

\section*{The Longitudinal Structure Function $F_L$}

\begin{figure}[t]
\includegraphics[clip,width=0.45\textwidth]{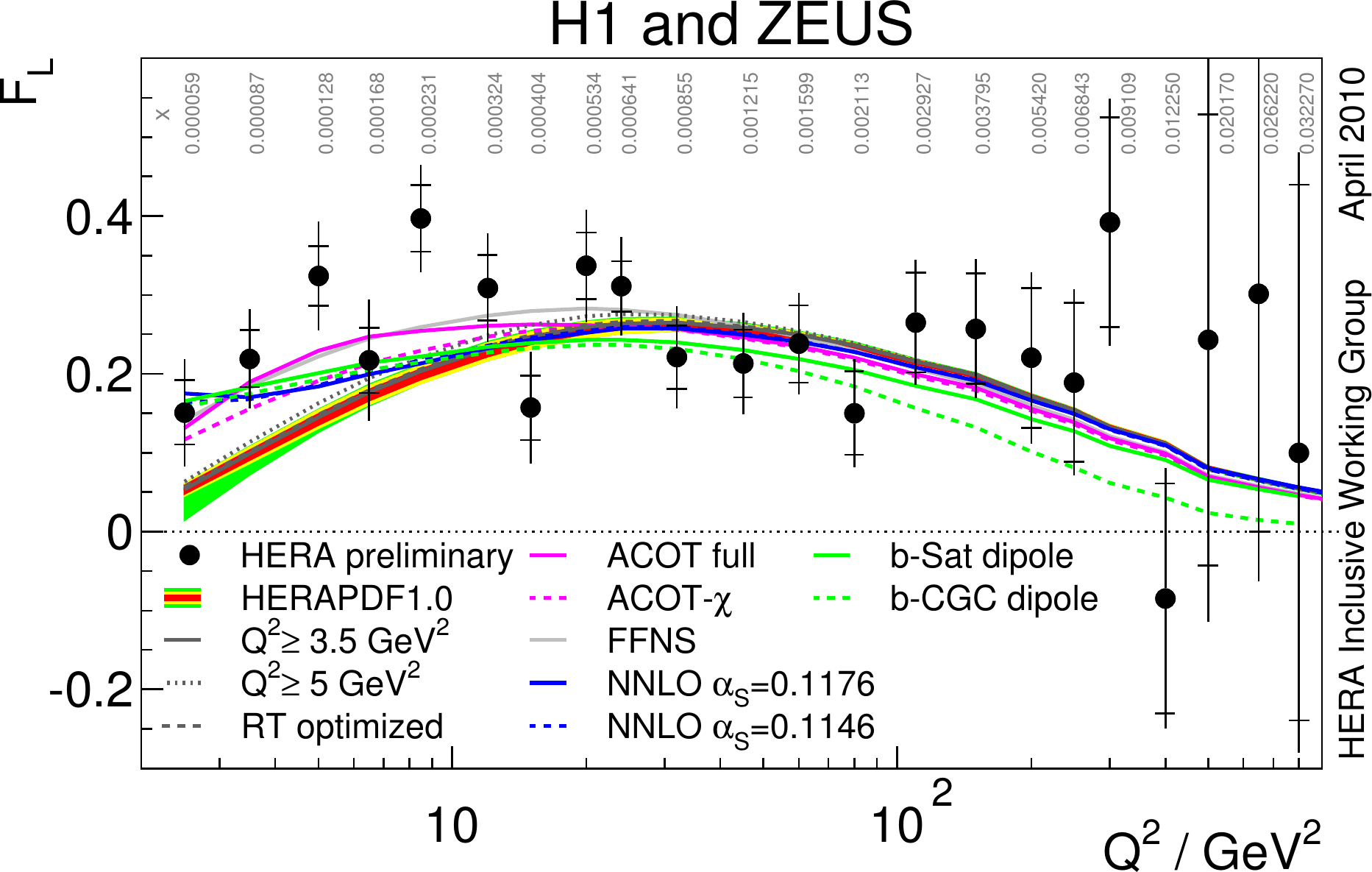}
\caption{Measurement of $F_L$ using the combined HERA 
data set from H1 and ZEUS. The data are compared with 
a selection of theoretical predictions. 
{\it (Figure  H1prelim-10-044 \& ZEUS-prel-10-008).}
\label{fig:FL}}
\end{figure}

\begin{figure*} 
\includegraphics[width=0.40\textwidth]{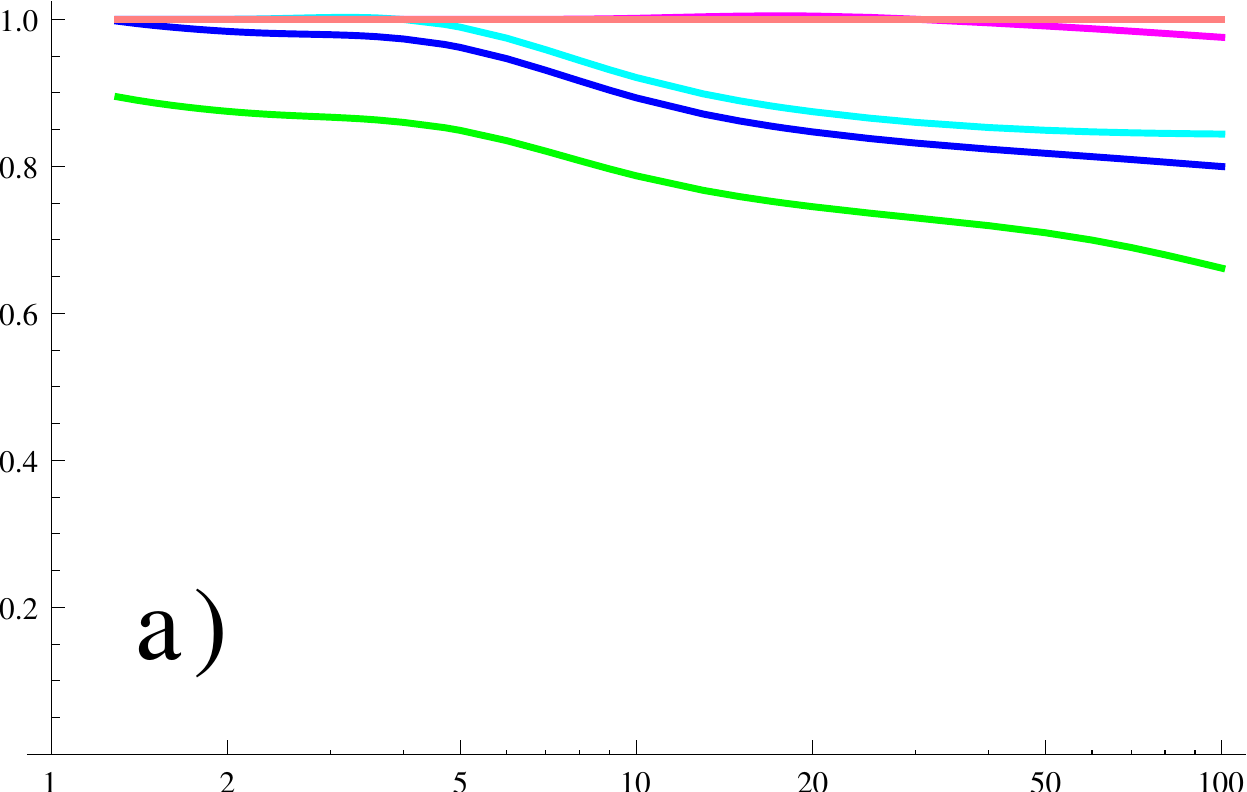}
\hfil
\includegraphics[width=0.40\textwidth]{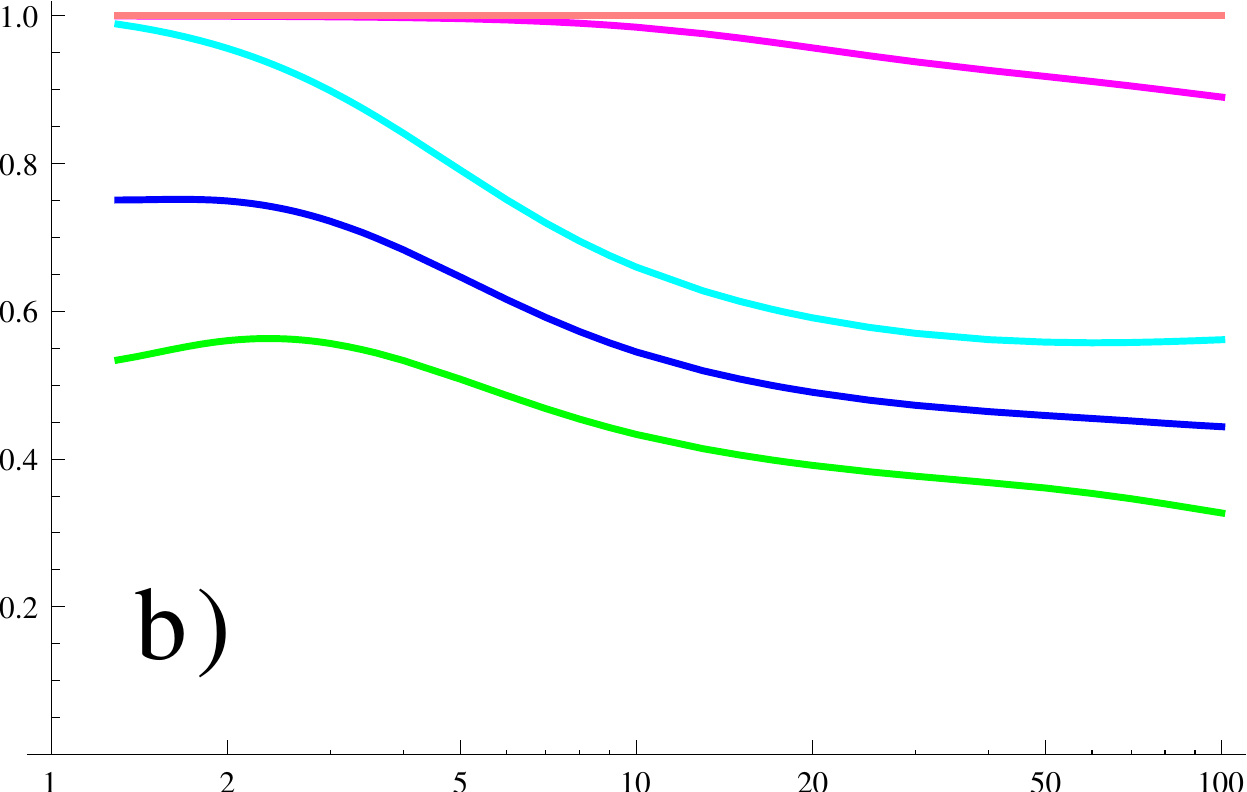}
\caption{Fractional flavor decomposition of  $F_L^i/F_L$ vs. $Q$
in GeV 
for  
a) $x=10^{-1}$ and  
b) $x=10^{-5}$.
Reading from the bottom, we plot 
the cumulative contributions for $\{u,d,s,c,b\}$.
\label{fig:FLrat}
}
\end{figure*} 

Most of the previous discussion has addressed the determination of 
the quark PDFs. Constraining the gluon PDF is a  challenge,
and the longitudinal structure function $F_L$ is particularly interesting 
as it involves both the heavy quark and the gluon distributions. 
Using the combined data from H1 and ZEUS, HERA has extracted 
 $F_L$ in an extended kinematic regime, and 
Figure~\ref{fig:FL} displays the result of the combined data as compared with 
various theoretical predictions. 

The measurement of $F_L$ is special for a number of reasons, and we write this  
schematically as:
 \begin{equation}
F_L \simeq 
\frac{m^2}{Q^2} \ q(x) 
+\alpha_S 
\left\{
C_g \otimes g(x) +
C_q \otimes q(x) 
\right\} \ .
\label{eq:FL}
 \end{equation}
Note that the LO term is zero in the limit of massless quarks as 
the $(m^2/Q^2)$ factor in 
Eq.~(\ref{eq:FL}) suppresses  the helicity violating contributions;
this is a consequence of the Callan-Gross relation.
Therefore, for light quarks the dominant contributions come from the NLO gluon term; hence, 
$F_L$ can provide useful information about the gluon PDF.

For the heavy quarks the picture is less obvious. While the NLO heavy quark contributions will 
clearly be small compared to the dominant gluon terms, the heavy quarks can contribute 
at LO if they can overcome the  $(m^2/Q^2)$ suppression. This is why the 
prediction of $F_L$ into the low $Q^2$ region as measured in 
Figure~\ref{fig:FL} is such a theoretical challenge. 
This raises  a number of questions: 
What is the flavor composition of $F_L$? 
Where are the heavy quark contributions important?

In Figure~\ref{fig:FLrat} we display
the fractional contributions to the structure functions
$F_L^i/F_L$ vs.  $Q$.
We observe that for
large $x$ and low $Q$ the heavy flavor contributions are minimal. For
example, in Figure~\ref{fig:FLrat}-a) at 
$Q \sim $ 5~GeV 
we see the  $u$-quark structure 
function $F_L^u$ 
comprises $\sim80\%$ of the total, $F_L^d$ is about 10\%, and
the $s$, $c$ and $b$ quarks divide the remaining fraction. 

At smaller $x$ values the picture changes and the heavy quarks 
are more prominent. 
In Figure~\ref{fig:FLrat}-b) for  $Q \sim 2 ~ GeV$ 
we see the  $u$-quark structure 
function $F_L^u$ 
comprises $\sim55\%$,  $F_L^d$ and $F_L^s$ are both  about 20\%, and
the $c$ and $b$ quarks make up the small remaining fraction. 
However,   $F_L^c$ increases quickly as $Q$ increases and 
is comparable to  $F_L^u$ ($\sim40\%$) for   $Q \sim 20 ~ GeV$.
Additionally, for  large $Q \sim 100 ~ GeV$ 
we see the contributions of the $u$-quark and $c$-quark
are comparable, the $d$-quark
and $s$-quark are comparable, and the relative sizes of the 
$u,c$\,  to $d,s$\,  terms are proportional to their couplings: 4/9 to 1/9. 
Thus, for low $x$ and intermediate to large $Q$ values 
we see that the quark masses (aside from the top) 
no longer play a prominent role and we 
approach the limit of \hbox{``flavor democracy.''}

\section*{Concluding Remarks}

We reviewed a number of recent developments regarding the extraction
and application of heavy quark Parton Distribution Functions (PDFs).
The high precision HERA measurements were essential in developing and
refining the theoretical treatment of the heavy quarks.  Even though
the accelerator facility stopped operation four years ago, the
analysis of the data continues.
The results of these analyses will provide the foundation upon which
future PDF analyses will be built, and the advances of the
experimental analysis and theoretical tools developed at HERA will
continue to influence future hadronic studies including those now
beginning at the LHC.

\subsection*{Acknowledgment}

We thank
M.~Botje,
A.~M.~Cooper-Sarkar,
A.~Glazov,
C.~Keppel,
J.~G.~Morf\'{\i}n,
P.~Nadolsky, 
J.~F.~Owens,
V.~A.~Radescu,
and
M.~Tzanov
for valuable discussions,
and we acknowledge the
hospitality of CERN, DESY, Fermilab, and Les Houches where
a portion of this work was performed.
F.I.O thanks the Galileo Galilei Institute for Theoretical Physics for
their hospitality and the INFN for partial
support during the completion of this work.
This work was partially supported
by the U.S.\ Department of Energy under grant DE-FG02-04ER41299,
and the Lightner-Sams Foundation.
The work of J.~Y.~Yu was supported
by the Deutsche Forschungsgemeinschaft (DFG) through grant No.~YU~118/1-1.
The work of K.~Kova\v{r}\'{\i}k was supported by the ANR projects
ANR-06-JCJC-0038-01 and ToolsDMColl, BLAN07-2-194882.
F.I.O. is grateful to DESY Hamburg and MPI Munich for their
organization and support of the Ringberg Workshop.



\nocite{*}
\bibliographystyle{elsarticle-num}
\bibliography{olness}

\end{document}